\documentstyle[psfig]{l-aa}

\begin{document}

   \thesaurus{07          
                           editors
              (08.01.3;   
               08.05.3;   
               08.16.4;   
               08.22.3;   
               08.13.2;   
               02.09.1)} 

   \title{The stability of 
late-type stars close to the Eddington limit}


   \author{Martin Asplund\inst{1,2}          }

   \offprints{martin@nordita.dk
              }

   \institute{Astronomiska observatoriet,
              Box 515,
              S--751 20 ~Uppsala,
              Sweden\\
              \and
              present address: NORDITA, 
              Blegdamsvej 17, 
              DK-2100 ~Copenhagen {\O}, 
              Denmark\\              
              }

   \date{Accepted for publication in {\it Astronomy \& Astrophysics}}

   \maketitle

   \markboth{M. Asplund: The stability of 
stellar atmospheres at the Eddington limit}
{M. Asplund: The stability of 
stellar atmospheres at the Eddington limit}

   \begin{abstract}

The opacity-modified Eddington limit has been computed for
hydrogen-deficient model atmospheres. The R Coronae Borealis (R\,CrB)
stars are found to be located strikingly 
close to the limit, which suggests that
the unknown trigger mechanism for their visual declines of the stars are
instabilities in connection with the stars
encountering the Eddington limit in their evolution. 
It also points to a similarity
between the eruptive behaviour of the R\,CrB stars and the
Luminous Blue Variables (LBVs).

Super-Eddington luminosities in hydrostatic model atmospheres 
manifest themselves by the presence of gas pressure inversions.
Such inversions are not an artifact of the assumption
of hydrostatic equilibrium but can also be present in 
hydrodynamical model atmospheres.
Only for very large mass loss rates hardly realized in supergiants
will the inversions be removed.
Instabilities may, however, still be present in such inversions, 
which is investigated 
for both H-rich and H-deficient late-type supergiant model atmospheres.
Dynamical instabilities may occur
in surface ionization zones, which might lead to ejections of gas.
A local, non-adiabatic, linear stability analysis reveals that 
sound waves can be amplified due to the strong radiative forces. 
However, despite the super-Eddington luminosities, 
the efficiency of the radiative instabilities is 
fairly low compared to for early-type stars  
with growth rates of $10^{-5}$\,s$^{-1}$. 

      \keywords{Stars: model atmospheres -- Instabilities -- 
Stars: AGB and post-AGB -- Stars: mass-loss --
Stars: evolution -- Stars: variables: R Coronae Borealis -- 
Stars: variables: Luminous Blue Variables              }
   \end{abstract}

\section{The Eddington limit and the development of 
gas pressure inversions}

The Eddington limit corresponds to a situation where the radiative 
acceleration outwards equals the gravitational 
acceleration inwards.
Eddington (1926) originally only considered electron scattering for
the opacity (the classical Eddington limit), but
in an ionization zone of an important element the true opacity can be
much higher.
Therefore, the opacity-modified Eddington limit in stellar 
atmospheres can
occur at significantly greater gravities than the classical Eddington
limit, as is evident for example from the calculations of Lamers \& 
Fitzpatrick (1988),
Gustafsson \& Plez (1992) and Asplund \& Gustafsson (1996).

If hydrostatic equilibrium is required, a stellar luminosity which
locally exceeds the Eddington luminosity automatically forces the
development of a gas pressure ($P_{\rm gas}$) inversion, 
as seen from the equation of hydrostatic equilibrium:

\begin{equation}
\frac{{\rm d}P_{\rm tot}}{{\rm dr}} = 
\frac{{\rm d}P_{\rm gas}}{{\rm dr}}
+ \frac{{\rm d}P_{\rm rad}}{{\rm dr}} = - g\rho.
\end{equation} 

\noindent From rearranging the above expression one obtains 

\begin{equation}
\frac{1}{\rho}\frac{{\rm d}P_{\rm gas}}{{\rm dr}} = - g_{\rm eff} 
= - g(1 - g_{\rm rad}/g) = - g\left[1 - \Gamma({\rm r})\right],
\end{equation}

\noindent with the radiative acceleration defined by:

\begin{equation}
g_{\rm rad} = - \frac{1}{\rho}\frac{{\rm d}P_{\rm rad}}{{\rm dr}} =
\frac{1}{c}\int_0^{+\infty}{\kappa_\nu F_\nu}d\nu 
\end{equation}

\noindent (e.g Mihalas 1978). 
Here $F_\nu$ denotes the physical flux and $\kappa_\nu$ 
the total mass extinction coefficient (with dimension cm$^{2}$\,g$^{-1}$). 
Obviously  a 
positive $P_{\rm gas}$-gradient must occur when the Eddington
limit is encountered if hydrostatic equilibrium is valid.

It should be emphasized that a density inversion does not
necessarily imply that the Eddington limit is exceeded. A density
inversion predominantly occurs due to a changing molecular weight
in the ionization zone of a dominant species, while a 
$P_{\rm gas}$-inversion reflects a super-Eddington luminosity due to 
high opacity (Asplund et al. 1997a). 
Since both are related to 
ionization they may 
often occur in the same atmospheric layers, 
which has probably caused the confusion 
found in the literature (e.g. Maeder 1989). 

It should be noted that
$P_{\rm gas}$-inversions only seem possible to develop in optically 
thick conditions. If the radiative acceleration exceeds gravity 
when lines dominate the opacity, 
the radiative force is highly unstable to 
perturbations induced by velocity gradients, since the spectral lines
may then absorb unattenuated continuum flux. 
Therefore, rather than the development of a 
$P_{\rm gas}$-inversion, a stellar wind will be initiated, 
which is the case for radiatively driven winds of
hot stars. The aim of the present study is to investigate whether 
$P_{\rm gas}$-inversions in late-type stars, where the opacity instead
is dominated by continuous opacity, are subject to a similar 
instability which would prevent their existence.
Before considering possible instabilities, in 
Sect. \ref{s:eddlimit} the derived Eddington limit for 
H-deficient stars is presented and a 
possible connection between the limit and the declines
of the R Coronae Borealis (R\,CrB) stars is discussed.

\section{The Eddington limit and the R\,CrB stars
\label{s:eddlimit}}

\subsection{The Eddington limit in the $T_{\rm eff}$-log\,$g$ diagram}

The location of the Eddington limit in a $T_{\rm eff}$-log\,$g$ 
diagram can be estimated by model atmosphere calculations
(Lamers \& Fitzpatrick 1988; Gustafsson \& Plez 1992). 
It should be emphasized that these models consistently include
the radiative acceleration in the
equation of hydrostatic equilibrium. Therefore, the location
of the limit is not obtained by extrapolation towards
lower gravities nor does it imply zero density, 
as has sometimes been claimed in the literature 
(e.g. Humphreys \& Davidson 1994;
Nieuwenhuijzen \& de Jager 1995).
An interesting conclusion from the above studies is that the
theoretical opacity-modified Eddington limit
coincides with the upper luminosity limit for stars
(Lamers \& Fitzpatrick 1988). It is therefore  
possible that the Eddington limit, possibly in connection 
with rotation (Langer 1997a,b; Owocki \& Gayley 1997), prevents 
the evolution of stars into the super-Eddington regime by
the existence of radiative instabilities.
These may give rise to the eruptions of
the Luminous Blue Variables (LBVs)
(cf. Humphreys \& Davidson 1994). 

Here a similar investigation but for the H-deficient 
R\,CrB and related stars will be carried out;
a preliminary report can be found in Asplund \& Gustafsson (1996).
The presence of $P_{\rm gas}$-inversions in such model 
atmospheres was noted by Asplund et al. (1997a), who speculated
that it may be the trigger
mechanism for the enigmatic visual declines of the 
R\,CrB stars. The variability of these stars is likely
due to the formation of an obscuring dust cloud close to the stellar
surface in the line-of-sight, but 
how the dust condensation can proceed despite the relatively high
equilibrium temperatures has long been an enigma (cf. Clayton 1996).
The presence of shocks seems to be a very promising mechanism for the
gas to cool sufficiently
and thereby trigger dust formation (Woitke et al. 1996).

\begin{figure}[t]
\centerline{
\psfig{figure=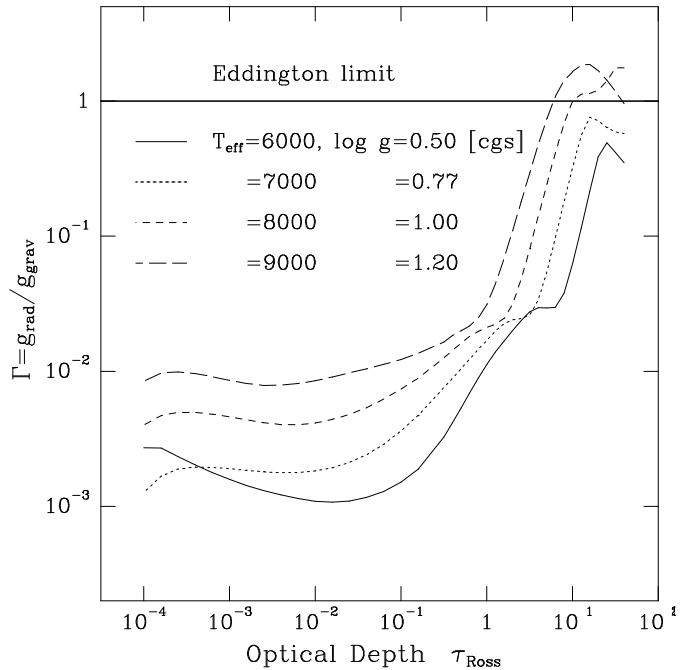,width=9.cm}}
   \caption{$\Gamma$
as a function of 
optical depth for a sequence of models with R\,CrB abundances
at constant luminosity and mass. All models have C/He = 1\%. 
The $T_{\rm eff} = 8\,000$\,K 
model is only partly affected by convection in the inner layers, 
while the hottest model is not at all. 
Increased molecular absorption causes the slight increase 
in $\Gamma$ towards the surface for the coolest model 
}
   \label{f:gamma}
\end{figure}

\begin{figure*}[t]
\centerline{
\psfig{figure=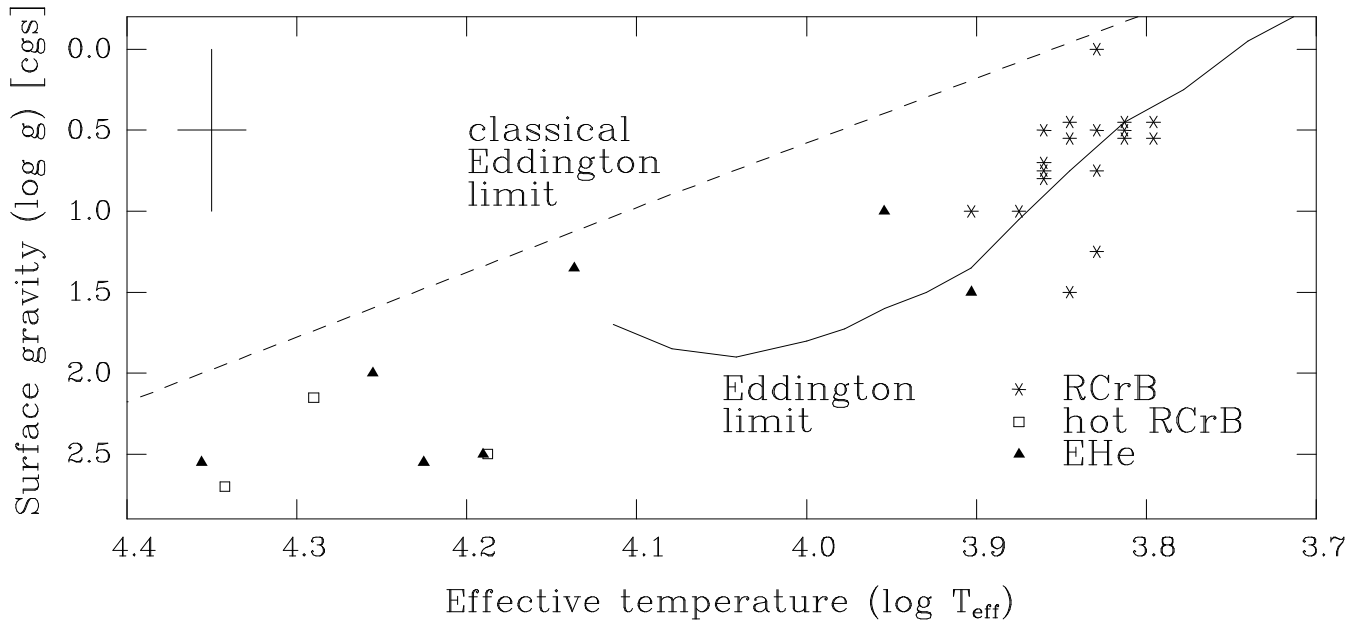,width=18cm}}
   \caption{The location of the opacity-modified Eddington limit 
(solid) in relation to the observed values of R\,CrB stars
(Lambert et al. 1997; Asplund et al. 1997c). Also 
shown is the classical Eddington limit (dashed) and 
the hot R\,CrB and EHe stars 
with determined parameters (Jeffery 1996). 
The cross in the upper left corner
illustrates the typical uncertainties in the 
estimated parameters of the R\,CrB stars 
}
   \label{f:eddlimit}
\end{figure*}

The models used here are recently constructed 
line-blanketed model photospheres with typical abundances of
R\,CrB stars (Asplund et al. 1997a).  
In Fig. \ref{f:gamma} the variation of $\Gamma$ with optical 
depth is displayed for a sequence of models corresponding to 
constant luminosity and mass 
(e.g. $L_*=8\cdot10^3\,$L$_{\sun}$ and $M_*=0.8\,$M$_{\sun}$). 
The super-Eddington luminosities
only occur in the deep atmospheric layers where He\,{\sc i} 
ionizes and, as expected, where $P_{\rm gas}$-inversions occur.
Since the R\,CrB stars presumably evolve towards higher $T_{\rm eff}$
(Kilkenny 1982), the stars will move from the sub-Eddington regime
into having super-Eddington luminosities, as is seen in
Fig. \ref{f:eddlimit} where the theoretical Eddington limit 
in the $T_{\rm eff}$-log\,$g$ diagram is shown. 
Compared with solar abundances the limit is shifted towards higher
$T_{\rm eff}$ due to the higher ionization potential of He
compared with H, but also
extends to slightly larger log\,$g$.
The location of the Eddington limit for higher 
$T_{\rm eff}$ has not yet been explored,
since a self-consistent hydrodynamical treatment is needed for 
$T_{\rm eff} \ga 13\,000$\,K when $\Gamma > 1$ occurs for 
$\tau_{\rm Ross} < 1$ and line opacity may determine $g_{\rm rad}$. 

The exact location of the computed Eddington limit will
naturally depend on the detailed properties of the
model atmospheres.
For $T_{\rm eff} \la 8\,500$\,K 
convection in the deeper layers of the He\,{\sc i} ionization zone 
restricts the limit to lower 
log\,$g$, since the radiative flux diminishes in the 
presence of a significant convective flux. 
Also, the lower temperatures decrease 
$\kappa_\nu$ and $g_{\rm rad}$ further. 
Including turbulent pressure would therefore  
bring the limit towards greater 
gravities by typically 0.2\,dex, due to less efficient convection with 
lower densities. 
A C/He ratio of 1\% has been assumed on the basis of the measured 
ratio in extreme helium (EHe) stars (Jeffery 1996), 
which presumably the R\,CrB stars 
are related to (Lambert et al. 1997). 
A higher carbon abundance pushes the Eddington limit 
towards higher gravities the same way as turbulent pressure does.
A lower overall metallicity by 0.5\,dex would decrease
the limit by about 0.2\,dex by lowering the temperatures in 
the inner layers with a diminished line-blanketing.

\subsection{Possible consequences for H-deficient stars}

In Fig. \ref{f:eddlimit} 
the estimated parameters of the R\,CrB stars
(Lambert et al. 1997; Asplund et al. 1997c) are also shown. 
Obviously, rather than being randomly located in the diagram 
as one a priori might have guessed, the R\,CrB stars fall 
close to the limit, which 
suggests that a connection between the 
declines of R\,CrB stars and the Eddington limit exists.
Thus, there is an alluring resemblance between the R\,CrB stars and the 
Luminous Blue Variables in their proximity to the Eddington
limits and their eruptive behaviour (Asplund \& Gustafsson 1996). 
If the R\,CrB stars are evolving towards higher $T_{\rm eff}$ 
(Kilkenny 1982) and 
the EHe domain at constant luminosity, 
their immediate progenitors should be the H-deficient 
carbon (HdC) stars (Lambert et al. 1997). 
The EHe and HdC stars experience no visual fadings despite being
otherwise similar to the R\,CrB stars. 
It is therefore plausible that the reason for the non-variability
is that the HdC stars have not yet 
reached the Eddington limit while the EHe stars are 
located on the stable side of the 
limit at higher $T_{\rm eff}$.

If the declines of the R\,CrB stars are indeed triggered by the
super-Eddington luminosities, one could expect
a correspondence between the maximum value of
$\Gamma$ in the atmosphere and frequency of declines.
In the pulsation-induced dust condensation model for the
R\,CrB stars (Woitke et al. 1996),
in which the dust formation is triggered by propagating shock
waves due to ordinary pulsations, one would sooner expect
a correlation with pulsation strength. Intriguingly, V854\,Cen, 
which is the R\,CrB star most frequent in decline, has very weak 
photospheric pulsations (Lawson \& Cottrell 1997) 
but by a wide margin the largest $\Gamma$ 
found of the 18 R\,CrB stars analysed sofar.
RY\,Sgr, on the other hand, has by far the strongest photospheric 
pulsations but only average amount of declines, which might 
reflect the fact that $\Gamma$ is not unusually high.
Pulsation may, however, push some stars across the limit by periodically
shifting them towards higher 
$T_{\rm eff}$, e.g. $\Delta T_{\rm eff} = \pm 250$\,K for R\,CrB 
itself (Rao \& Lambert 1997).

The proposed connection between the Eddington limit and the declines
of the R\,CrB stars is certainly not without problems,
which deserve to be addressed further.
A few stars seem not to behave as expected if the proposal
is correct, in particular this is true for some of the
coolest EHe stars and the three hot R\,CrB stars.

\section{The disappearance of gas pressure inversions
\label{s:pgasinversion}}

\subsection{Effects of convection}

As seen above, a $P_{\rm gas}$-inversion must develop
at the Eddington limit in hydrostatic equilibrium.
One should ask, however, whether
strong convection may take place to prevent it, since
the ionization zone
causing the high opacity and thereby the super-Eddington
luminosities also favours convection. 
In fact, a super-Eddington luminosity in the diffusion regime
automatically requires the presence of convection as shown 
by Langer (1997a).
Generalizing Langer's derivation for a non-negligible
radiation pressure to also include the
effects of ionization, the Schwarzschild criterion 
$\nabla_{\rm rad} > \nabla_{\rm ad}$ for the onset of
convection is equivalent to 

\begin{equation}
\Gamma({\rm r}) > 
\frac{4\alpha
\left\{ 2 + 8\alpha + x\left(1-x\right)
\left[5/2 + \epsilon/kT + 4\alpha\right]
\right\} }
{5 + 40\alpha + 32\alpha^2 + x\left(1-x\right)
\left[5/2 + \epsilon/kT + 4\alpha\right]^2}
\end{equation}

\noindent 
(e.g. Mihalas \& Weibel Mihalas 1984).
Here \mbox{$\alpha=P_{\rm rad}/P_{\rm gas}$}, while $x$ denotes the
ionization fraction of the most abundant element with an
ionization potential $\epsilon$.
Convection must be initiated as $\Gamma$(r) approaches unity, 
since the right hand side is always less than 1.
Thereby the radiative flux and hence $\Gamma$ diminishes
when convection is efficient.
The Eddington limit can therefore not be exceeded 
in adiabatic convection zones in the stellar
interior and hence $P_{\rm gas}$-inversions will not develop. 
Thus, inversions can only be present 
in non-adiabatic convection zones close to the stellar surface.

It should be remembered that the 
location of the Eddington limit is
dependent on the convection treatment.
A $P_{\rm gas}$-inversion will for example disappear 
if the density scale height 
is applied instead of the normal pressure scale height 
for the calculation of the
convective flux through the mixing length theory. 
The mixing length theory is rarely a good
description of stellar convection, and in particular for these
extreme situations it may be very misleading.
Ultimately, hydrodynamical simulations will be necessary to
investigate the effects of convection on $P_{\rm gas}$-inversions.

\subsection{Effects of mass loss
\label{s:massloss}}

Using the momentum equation one can 
investigate under which conditions a $P_{\rm gas}$-inversion can exist
in the presence of a stellar mass loss. Or equivalently: which
mass loss rates can one expect from a star if the super-Eddington
luminosities would instead drive a stellar wind?

In the presence of a stellar outflow, 
the equation of hydrostatic equilibrium 
must be replaced with the momentum equation:

\begin{equation}\label{e:momentumeq}
 v\frac{{\rm d}v}{{\rm dr}} =
- \frac{1}{\rho}\frac{{\rm d}P_{\rm gas}}{{\rm dr}}
 - g_{\rm eff},
\end{equation} 

\noindent 
where we have restricted ourselves to the one-dimensional, 
time-independent case.
Also other external forces, such as turbulent forces 
($g_{\rm turb} = - 1/\rho \cdot {\rm d}P_{\rm turb}/{\rm dr}$)
and centrifugal forces 
($g_{\rm cent} = v_{\rm rot}^2(r,\theta)/r$),
can be contained in $g_{\rm eff}$ besides radiation. 
It should therefore be kept in mind 
that the possibility of an instability may be
underestimated when restricting the following discussion
to only radiative forces, as other
forces also tend to be destabilizing
(e.g. Nieuwenhuijzen \& de Jager 1995;
Langer 1997a,b; Owocki \& Gayley 1997).

Together with a constant,
spherically symmetric mass loss rate $\dot M = 4\pi r^2 \rho v$,
and the equation of state 
\mbox{$P_{\rm gas} = {\cal R} \rho T / \mu$}, 
it is possible to
rewrite the momentum equation into the so called ``wind equation":

\begin{equation}
\left(v^2 - c_{\rm T}^2\right)
\frac{1}{v}\frac{{\rm d}v}{{\rm dr}} = 
c_{\rm T}^2\left(\frac{2}{r} - 
\frac{1}{T}\frac{{\rm d}T}{{\rm dr}} +
\frac{1}{\mu}\frac{{\rm d}\mu}{{\rm dr}}\right)
- g_{\rm eff}, 
\label{e:windeq}
\end{equation} 

\noindent where the isothermal sound speed 
defined by 
$c_{\rm T} ^2 = 
(\partial P_{\rm gas}/\partial \rho)_{\rm T} = 
P_{\rm gas}/\rho$ has been introduced.
By an inspection of Eq. \ref{e:windeq}, it is 
possible to determine under which circumstances $\rho$- or 
$P_{\rm gas}$-inversions are allowed.
In normal cases, d$\mu$/dr$\ge0$ in the surface layers
and therefore the parenthesis on the right hand side
of Eq. \ref{e:windeq} is positive as long as a temperature
inversion is not present, which is not possible in
optically thick layers with a radiative flux. 
Therefore, if $g_{\rm rad} > g$ 
but $v < c_{\rm T}$, the velocity gradient 
${\rm d}v/{\rm dr}$ must be negative. Since  
a constant $\dot M$ has been assumed, 
a density inversion will be present
if $r/v \cdot {\rm d}v/{\rm dr} + 2 < 0$
and therefore a $P_{\rm gas}$-inversion is possible 
(though not required). 
If, on the other hand, $v > c_{\rm T}$ while 
$g_{\rm rad} > g$, then ${\rm d}v/{\rm dr}$
must be positive, and the density gradient
cannot be positive: a density inversion cannot exist. Since a 
$P_{\rm gas}$-inversion requires a density inversion, neither inversion 
can occur in the stellar atmosphere despite the super-Eddington 
luminosities. Again the possibility
of a temperature inversion has been neglected.
A sufficient criterion for the disappearance of a 
$P_{\rm gas}$-inversion is therefore that the 
outflow velocity due to mass loss 
is high enough ($v \ge c_{\rm T}$).

The critical mass loss rate can be estimated by comparing
$v = \dot M / 4\pi r^2 \rho$ with $c_{\rm T}$.
The values for $c_{\rm T}$ and $\rho$ 
will be taken from static models, which is justified for 
these rough estimates, since the atmosphere
below the sonic point is normally little affected by mass loss.
A model atmosphere 
with $T_{\rm eff}=7\,000$\,K, log$\,g=0.5$ [cgs] and solar 
abundances requires a critical mass loss rate 
$\dot M_{\rm crit} \approx 
6 \cdot 10^{-4} (M_*/M_{\sun})\,{\rm M}_{\sun}\,{\rm yr}^{-1}$.
Hence $P_{\rm gas}$-inversions can exist in stars
despite high mass loss rates.
Achmad \& Lamers (1997) have arrived at similar conclusions 
and estimates. 
They also verify the finding numerically with 
dynamical steady state atmospheres, 
though not altogether self-consistently.

An even higher critical mass loss rate is found for the
R\,CrB stars. The lack of hydrogen
makes the continuous opacity significantly lower than for solar
abundances and hence the density at a given optical depth will
be correspondingly greater, typically by a factor of 50 
(Asplund et al. 1997a).
Since the density is higher, the outflow velocity 
will be lower in order to carry the same $\dot M$. A similar
model as above but with abundances typical for R\,CrB stars, 
will therefore require $\dot M_{\rm crit} \approx 
3 \cdot 10^{-2} (M_*/M_{\sun})\,{\rm M}_{\sun}\,{\rm yr}^{-1}$. 
The mass loss rates of R\,CrB stars are poorly known, 
in particular since it may be episodic, but rough
estimates suggest
\mbox{$\dot M \sim 10^{-6}\,{\rm M}_{\sun}\,{\rm yr}^{-1}$}
(e.g. Feast 1986), orders of magnitude smaller than $\dot M_{\rm crit}$.

It is concluded that $P_{\rm gas}$-inversions are not an 
artifact of hydrostatic equilibrium but may also exist in 
dynamical steady state atmospheres, even with 
a high mass loss rate.


\section{Dynamical instabilities
\label{s:dynamical}}


Dynamical stability has been extensively studied, in particular
in connection with the termination of the AGB 
(e.g. Paczy\'{n}ski \& Zi\'{o}{\l}kowski 1968; 
Wagenhuber \& Weiss 1994).
Radial adiabatic oscillations will grow exponentially in time
for dynamical instability,
which is formally encountered if the 
first generalized adiabatic exponent 

\begin{equation}\label{e:gamma1}
\Gamma_1 = 
\left(\frac{\partial {\rm ln}P_{\rm tot}}{\partial 
{\rm ln}\rho}\right)_{\rm ad}
\end{equation} 

\noindent 
is less than $4/3$ in a region effectively
isolated from the rest of the star. An ideal gas has $\Gamma_1 = 5/3$
while pure radiation has $\Gamma_1 = 4/3$, but the combined effect
of non-negligible radiation pressure and ionization may 
push $\Gamma_1$ below the limiting value. 
For dynamical instability to occur, the pressure weighted 
volumetric mean value of $\Gamma_1$ between the depth $r$ and
the stellar surface $R_*$ 

\begin{equation}\label{e:gamma1}
\langle \Gamma_1 \rangle = \int^{R_*}_r \Gamma_1 P_{\rm tot} {\rm d}V \bigg/ 
\int^{R_*}_r P_{\rm tot} {\rm d}V
\end{equation} 

\noindent must be reduced below $4/3$
in a significant fraction of the star.
Stars with a pronounced core-envelope structure and high 
luminosity-to-mass ratios, like the supergiants investigated here, 
may show dynamical instabilities in ionization zones
(e.g. Stothers \& Chin 1993; Wagenhuber \& Weiss 1994).

\begin{figure}[t]
\centerline{
\psfig{figure=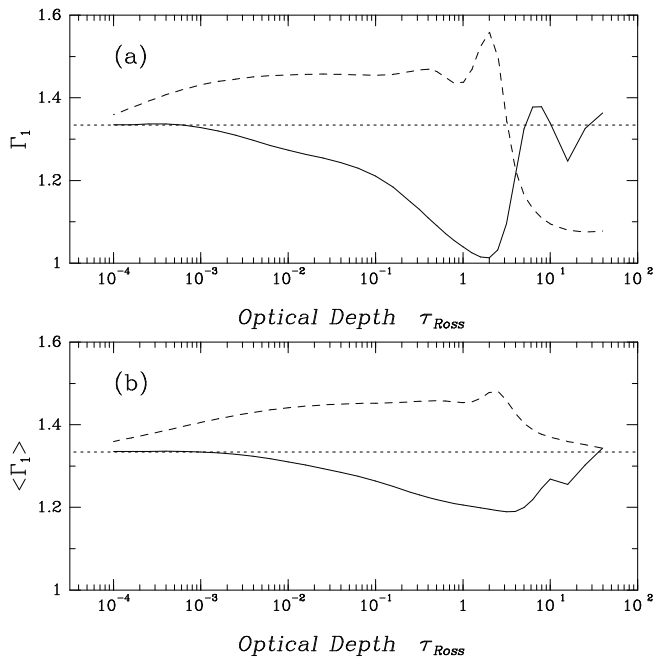,width=9.0cm}}
   \caption{{\bf a} The generalized first adiabatic exponent
$\Gamma_1$ as a function of optical depth in model atmospheres
with solar (solid) and R\,CrB (dashed) compositions.
{\bf b} The pressure weighted first 
adiabatic exponent $\langle \Gamma_1 \rangle$ for the same models as in
{\bf a}. In both panels the 
parameters are $T_{\rm eff} = 7\,000$ and log\,$g = 0.50$.
Values below the dotted lines at $\Gamma_1 = \langle \Gamma_1 \rangle = 4/3$ 
correspond formally to dynamical instability}
   \label{f:gamma1}
\end{figure}

In Fig. \ref{f:gamma1} the run of $\Gamma_1$ and 
$\langle \Gamma_1 \rangle$ as 
functions of optical depth are shown for two model atmospheres
corresponding to $T_{\rm eff} = 7\,000$ and log\,$g = 0.50$
for a solar and a H-deficient composition. 
In particular, $\langle \Gamma_1 \rangle$ is
reduced much below $4/3$  for the model with solar abundances
in the H\,{\sc i} and He\,{\sc i} ionization zones. 
At the surface $P_{\rm rad}$ dominates the total pressure
and hence $\Gamma_1$ is very close to $4/3$. 
A more complete study for H-rich
model atmospheres has been presented by Lobel et al. (1992), who
arrive at the same conclusions.
In the R\,CrB model only the He\,{\sc i} ionization zone exists, 
which explains the quite different
depth variation of $\Gamma_1$.
The higher temperatures at depth in
the H-deficient model make He\,{\sc i} ionization occur 
at smaller $\tau_{\rm Ross}$.
Also seen is the minor effect of the C\,{\sc i} 
ionization zone at $\tau_{\rm Ross} \simeq 1$. 
These models only extend down to $\tau_{\rm Ross} = 40$ 
and hence it is not possibly to tell how 
$\langle \Gamma_1 \rangle$ varies
further in, though the He\,{\sc ii} ionization zone  
will also reduce $\Gamma_1$. 

It seems that atmospheres of late-type 
supergiants are close to being dynamically unstable, or may
even be so for sufficiently high luminosity-to-mass ratios.
Violent instabilities due to this might lead to ejection of material,
as found for the termination of the AGB 
(Wagenhuber \& Weiss 1994), as well as for the LBVs 
(Stothers \& Chin 1993).
Also R\,CrB stellar models seem to suffer from a similar
dynamical instability as the stars on the tip of the AGB
(A. Weiss, private communication).
It should be remembered though that the above-mentioned 
evolutionary studies lack a proper hydrodynamical treatment, 
and therefore interpreting the instabilities found in
the models as actually occurring in the stars must still be made
with some caution. 

\section{Radiative instabilities}

\subsection{Radiatively driven sound waves in 
$P_{\rm gas}$-inversions
\label{s:radinstab}}

It is clear from the discussion in 
Sect. \ref{s:pgasinversion} and \ref{s:dynamical} 
that $P_{\rm gas}$-inversions may exist in stars with high
luminosity-to-mass ratios and that such stars are close
to being dynamically unstable. Thus, these stars may 
be vulnerable to radiative instabilities, since the additional
destabilization from radiative forces may render them unstable. 

The existence of radiatively driven sound
waves in hot stars due to high $g_{\rm rad}$ has been
suggested by Hearn (1972, 1973),
and further elaborated by Berthomieu et al. (1976), 
Spiegel (1976) and Carlberg (1980).
The instability is based on the increase in opacity upon compression,
since the radiation force is in phase with 
the velocity perturbation and hence causes amplification.
For these high temperatures it requires essentially 
isothermal perturbations, since 

\begin{equation}
(\partial\kappa_{\rm F}/\partial \rho$)$_T + 
(\Gamma_3-1)(\partial\kappa_{\rm F}/\partial T$)$_\rho < 0.
\end{equation} 

\begin{figure}[t]
\centerline{
\psfig{figure=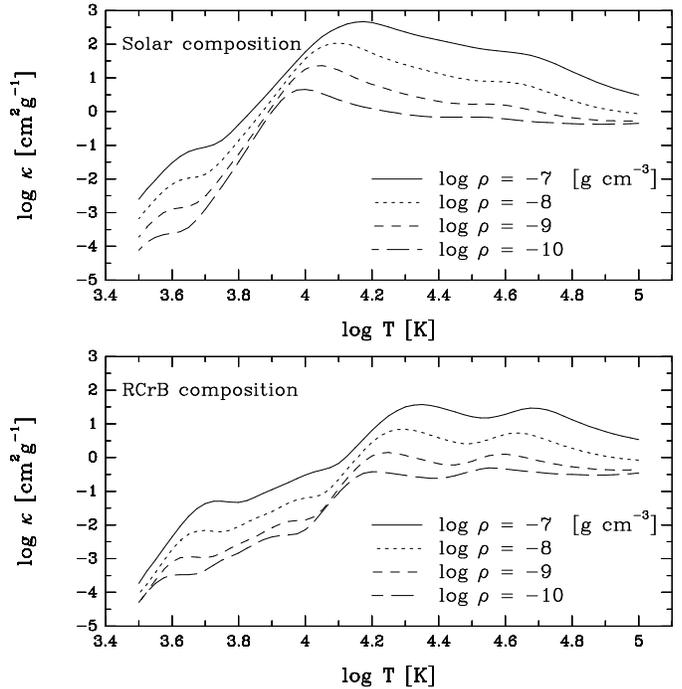,width=9.5cm}}
   \caption{The dependence of the Rosseland mean opacity 
on temperature and density
for {\bf a} solar abundances and {\bf b} R\,CrB abundances.
The atomic data for the opacities are taken from the 
Opacity Project (Seaton et al. 1994 and references therein).
The top of the $P_{\rm gas}$-inversion for a model atmosphere
with $T_{\rm eff} = 7\,000\,$K and log\,$g = 0.5$
occurs at log\,$T \approx 3.9$ for solar abundances and at 
log\,$T \approx 4.1$ for an R\,CrB composition
}
   \label{f:opacity}
\end{figure}

\noindent
Here $\Gamma_3 -1 = 
(\partial {\rm ln} T/ \partial {\rm ln} \rho)_{\rm ad}$, while
$\kappa_{\rm F}$ is the flux weighted mean opacity. 
Thus only short wavenumbers will be amplified. 
The growth rate is especially high as
$\Gamma$ approaches unity (Berthomieu et al. 1976) but 
the super-Eddington case has not been investigated. 
Carlberg (1980) found that the instability may
provide rapid transfer of momentum from the 
radiation to the gas with typical growth rates of 
$10^{-3}$\,s$^{-1}$ for $T_{\rm eff} \sim 50\,000$\,K.
It is possible that a similar instability is
exhibited also in late-type supergiants, with the 
important differences being the super-Eddington luminosities 
in the $P_{\rm gas}$-inversions and that isothermal 
perturbations are not necessary, since both 
($\partial\kappa_{\rm F}/\partial\rho$)$_{\rm T} > 0$ 
and ($\partial\kappa_{\rm F}/\partial T$)$_\rho > 0$ 
in the upper layers of the $P_{\rm gas}$-inversions
(Fig. \ref{f:opacity}). 

To study the stability of a $P_{\rm gas}$-inversion a local, linear,
non-adiabatic stability analysis has been carried out.
Small, sinusoidal 
perturbations around the equilibrium values are assumed
for the temperature, density and vertical velocity:

\begin{eqnarray*}
T & = & T_0 + T_1 \cdot {\rm e}^{i(kr - \omega t)}  \\
\rho & = & \rho_0 + \rho_1 \cdot {\rm e}^{i(kr - \omega t)}  \\
v & = & v_0 + v_1 \cdot {\rm e}^{i(kr - \omega t)},   
\end{eqnarray*}

\noindent 
where $T_1, \rho_1$ and $v_1$ are small.
The atmosphere is assumed to be static ($v_0=0$), since the outflow
velocity must anyway be less than the isothermal sound speed for a 
$P_{\rm gas}$-inversion to exist, as demonstrated above. 
The analysis will be restricted to vertical motion
and therefore multi-dimensional instabilities of 
Rayleigh-Taylor-type will not be found. 
Carlberg (1980) investigated also such a gradient instability
for hot stars but found slow growth rates compared to the radiatively
driven sound waves.
Such a gradient instability would resemble 
convective motion which, however,
is already present in the layers of interest in the
unperturbed model and is thus not specifically searched for here. 

The equations to be
linearized are the material
equations for conservation of
mass, momentum and energy:

\begin{eqnarray}
\frac{{\rm D}\rho}{{\rm Dt}} + 
 \rho \left(\nabla \cdot {\bf v}\right) & = & 0  
\label{e:hydroeq1}
\\
\frac{{\rm D{\bf v}}}{{\rm Dt}} +
\frac{1}{\rho} \nabla P_{\rm gas}
& = &
- g_{\rm eff}\hat{{\rm r}} 
\label{e:hydroeq2}
\\
\rho\frac{{\rm D} e}{{\rm Dt}} -
\frac{P_{\rm gas}}{\rho} 
\frac{{\rm D\rho}}{{\rm Dt}} & = &  
4\pi\rho\kappa_{\rm Ross}^{\rm abs}
\left( J - B \right) 
- \rho {\bf v} \cdot g_{\rm rad}\hat{{\rm r}} 
\label{e:hydroeq3}
\end{eqnarray}

\noindent (e.g. Mihalas \& Weibel Mihalas 1984), where D/Dt,
$e$, $B$ and $J$ denote the Lagrangean co-moving derivative,
the specific internal energy of the gas,
the wavelength integrated
Planck function ($B = \int^\infty_0 B_\nu(T){\rm d}\nu$) and
mean intensity ($J = \int^\infty_0 J_\nu{\rm d}\nu$), 
respectively. 
The absorption and Planck mean opacities have been
replaced by the Rosseland mean opacity for absorption
$\kappa_{\rm Ross}^{\rm abs} = \kappa_{\rm Ross} - \sigma$,
which is a good approximation 
for the relevant optical depths. 
The scattering contribution
to the total extinction is here denoted by $\sigma$.
Furthermore LTE ($S_\nu = B_\nu$) has been assumed.

The material equations must be coupled to the equations
for the radiation field
to account for radiative transfer
effects through exchanges of energy and momentum between
the photons and the gas. 
In this exploratory investigation the effects 
of radiation are included with a time-independent treatment,
i.e. the radiation field is assumed to be quasi-static. 
The radiation momentum and energy equations will then look like:

\begin{eqnarray}
\frac{{\rm d} K}{{\rm dr}} & = & - \rho \kappa_{\rm F} H \\
\frac{{\rm d} H}{{\rm dr}} & = &
\rho \left(\kappa_{\rm Ross} - \sigma\right) 
\left(B - J \right).
\end{eqnarray}

\noindent Here 
$H$ and $K$ denote the wavelength integrated
Eddington flux ($H = (1/4\pi)\int^\infty_0 F_\nu{\rm d}\nu$)
and radiation pressure ($K = (c/4\pi) P_{\rm rad}$)
respectively. 
In the diffusion regime for the unperturbed atmosphere,
$\kappa_{\rm F} =\kappa_{\rm Ross}$ is valid. 
The equilibrium radiation quantities are related by:

\begin{equation}  
J_0 = B_0 = 4 (T_0/T_{\rm eff})^4 H_0 = 
\sigma_{\rm R} T_0^4 / \pi.
\end{equation}

\noindent 
For the first relation radiative equilibrium is implicitly
imposed, which is not strictly true in the cases which
will be investigated below. However, the analysis will be
restricted only to atmospheric depths
where the convective flux carries $< 10\%$ of the total flux 
(i.e. in the upper part of the convection zone) and radiative
equilibrium can therefore be largely justified.
To close the relations the Eddington
approximation ($K=J/3$) will be made, which is, like
LTE,  well justified
in the atmospheric layers of interest.
In the following, the radiation energy and 
momentum equations will be combined
to a single equation, which means that there is one more 
variable to perturb, for which $J$ is chosen:

\begin{eqnarray*}
J & = & J_0 + J_1 \cdot {\rm e}^{i(kr - \omega t)}.   
\end{eqnarray*}

Together with the definitions of 
the isothermal sound speed $c_{\rm T}$, the density scale height
$H_\rho = - [{\rm d ln}\rho/{\rm dr}]^{-1}$ and an ideal gas for $P_{\rm gas}$, 
linearization 
to first order yields:

\begin{eqnarray}
\frac{\rho_1}{\rho_0}\left[-i\omega\right] +
v_1\left[ik - H_\rho^{-1}\right] & = & 0 \\
\frac{\rho_1}{\rho_0}\left[g +
c_{\rm T}^2\left(ik + \frac{1}{T_0}\frac{{\rm d}T}{{\rm dr}}
 - \frac{1}{\mu_0}\frac{{\rm d}\mu}{{\rm dr}}
\right)\right] 
& + & \nonumber \\
+ \frac{\rho_1}{\rho_0}\left[- c_{\rm T}^2
\mu_\rho
\left(ik - H_\rho^{-1} +
\frac{1}{T_0}\frac{{\rm d}T}{{\rm dr}} -
\frac{2}{\mu_0}\frac{{\rm d}\mu}{{\rm dr}}\right)\right] 
& + & \nonumber \\
+ \frac{T_1}{T_0}\left[
c_{\rm T}^2\left(ik - H_\rho^{-1}
 - \frac{1}{\mu_0}\frac{{\rm d}\mu}{{\rm dr}}
\right)\right] 
& + & \nonumber \\
+ \frac{T_1}{T_0}\left[- c_{\rm T}^2
\mu_T
\left(ik - H_\rho^{-1} +
\frac{1}{T_0}\frac{{\rm d}T}{{\rm dr}} -
\frac{2}{\mu_0}\frac{{\rm d}\mu}{{\rm dr}}\right)\right] 
& + & \nonumber \\
+ v_1\left[ - i\omega\right] 
+ \frac{J_1}{J_0}\left[- ik\frac{4\sigma_{\rm R} T_0^4}{3c\rho_0} \right]
& = & 0 \\
\frac{\rho_1}{\rho_0}\left[-i\omega \left(\rho_0
\frac{\partial e}{\partial \rho} 
- c_{\rm T}^2 \right)\right] 
& + & \nonumber \\
+ \frac{T_1}{T_0}\left[-i\omega T_0 \frac{\partial e}{\partial T} +
16\left(\kappa_0 - \sigma_0\right)\sigma_{\rm R} T_0^4\right]
& + & \nonumber \\
+ v_1\left[g_{\rm rad} 
+ \frac{{\rm d}e}{{\rm dr}} + 
c_{\rm T}^2 H_\rho^{-1}  \right]
& + & \nonumber \\
+ \frac{J_1}{J_0}\left[-4\left(
\kappa_0 - \sigma_0\right)\sigma_{\rm R} T_0^4\right]
& = & 0  \\
\frac{\rho_1}{\rho_0}\left[-\frac{ik}{\rho_0(\kappa_0 - \sigma_0)}
\frac{1}{4}\left(\frac{T_{\rm eff}}{T_0}\right)^4\left(1 + 
\frac{\partial {\rm ln} \kappa}{\partial {\rm ln} \rho}\right)\right] 
& + & \nonumber \\
+ \frac{T_1}{T_0}\left[-4 - \frac{ik}{\rho_0(\kappa_0 - \sigma_0)}
\frac{1}{4}\left(\frac{T_{\rm eff}}{T_0}\right)^4
\frac{\partial {\rm ln} \kappa}{\partial {\rm ln} T}\right]
& + & \nonumber \\
+ \frac{J_1}{J_0}\left[1 + 
\frac{k^2}{3\kappa_0(\kappa_0 - \sigma_0)\rho_0^2}\right]
& = & 0.  
\end{eqnarray}

\noindent 
The notation has been abbreviated according to 
\mbox{$\mu_\rho = (\rho_0/\mu_0)(\partial \mu/\partial \rho)$}
and \mbox{$\mu_T = (T_0/\mu_0)(\partial \mu/\partial T)$}.
Derivatives with respect to $v$ and $J$ have been ignored,
which is justified in optically thick layers where
line opacity is unimportant. True enough, photoionization
edges introduce some velocity dependence on $g_{\rm rad}$
but this minor effect is neglected. 

\begin{table}[t]
\caption{Local values of the atmospheric structure. 
Both the model with a solar and an R\,CrB composition
have $T_{\rm eff} = 7000$ and log\,$g = 0.5$ 
\label{t:values}
}
\begin{tabular}{lccc} 
\\
 \hline \\
Variable &  Units & Solar  &  R\,CrB \\
\hline \\ 
log\,$\tau_{\rm Ross}$ &  & 0.5 & 1.2 \\
$g_{\rm rad}$ & [cm\,s$^{-2}$] &  $1.29\cdot10^1$ & $4.31\cdot10^0$\\
$T$ & [K] &  9610 & 14950 \\
$\rho$ &  [g\,cm$^{-3}$] & $4.62\cdot10^{-11}$ & $4.96\cdot10^{-9}$ \\
$P_{\rm gas}$ & [dyn\,cm$^{-2}$] & $5.19\cdot10^1$ & $1.92\cdot10^3$ \\
$c_{\rm s}$ & [cm\,s$^{-1}$] & $1.32\cdot10^6$ & $6.67\cdot10^5$ \\
$(1/T)$d$T$/dr & [cm$^{-1}$] & $-6.67\cdot10^{-12}$ & $-3.15\cdot10^{-11}$ \\
$(1/\rho)$d$\rho$/dr & [cm$^{-1}$] & $2.10\cdot10^{-11}$ & $9.87\cdot10^{-11}$ \\
$(1/\mu)$d$\mu$/dr & [cm$^{-1}$] & $5.92\cdot10^{-12}$ & $6.50\cdot10^{-11}$ \\
$\kappa_{\rm Ross}$ & [cm$^2$\,g$^{-1}$] & $2.50\cdot10^0$ & $9.11\cdot10^{-1}$ \\
$\sigma$ & [cm$^2$\,g$^{-1}$] & $2.55\cdot10^{-1}$ & $2.94\cdot10^{-2}$ \\
$\partial{\rm ln}\kappa/\partial{\rm ln}T$ &  & 
$2.43\cdot10^{-1}$ & $1.32\cdot10^{1}$ \\
$\partial{\rm ln}\kappa/\partial{\rm ln}\rho$ & & 
$6.75\cdot10^{-1}$ & $1.72\cdot10^{-1}$ \\
d$e$/dr & [cm$^{1}$\,s$^{-2}$] & $-1.29\cdot10^{2}$ & $-5.15\cdot10^{2}$ \\
$\rho (\partial e/\partial\rho)_T$ & 
[cm$^{2}$\,s$^{-2}$] & $-2.15\cdot10^{12}$ & $-8.21\cdot10^{11}$ \\
$T (\partial e/\partial T)_\rho$ & 
[cm$^{2}$\,s$^{-2}$] & $2.14\cdot10^{13}$ & $1.63\cdot10^{13}$ \\
\hline  \\
\end{tabular}

\end{table}

In order for the above set of equations to have a non-trivial 
solution for the perturbations $T_1, \rho_1$, $v_1$ and $J_1$, the
determinant of the equations must be equal to 0. 
The three
complex eigenvalues $w$ for a given real wavenumber $k$ 
then correspond to the allowed perturbations. 
Two of the modes will represent radiation-modified
gravity-acoustic waves while the third is a 
non-propagating, heavily damped, thermal wave.
Here two different cases will be investigated numerically, one
corresponding to a $P_{\rm gas}$-inversion in a supergiant with
solar abundances and the second the same but for an R\,CrB star.
As before, both model atmospheres have 
$T_{\rm eff} = 7\,000\,$K and log$\,g = 0.5$.
Values for some of the physical parameters at some atmospheric
layer where the instability may occur 
are found in Table \ref{t:values}.

\begin{figure}[t]
\centerline{
\psfig{figure=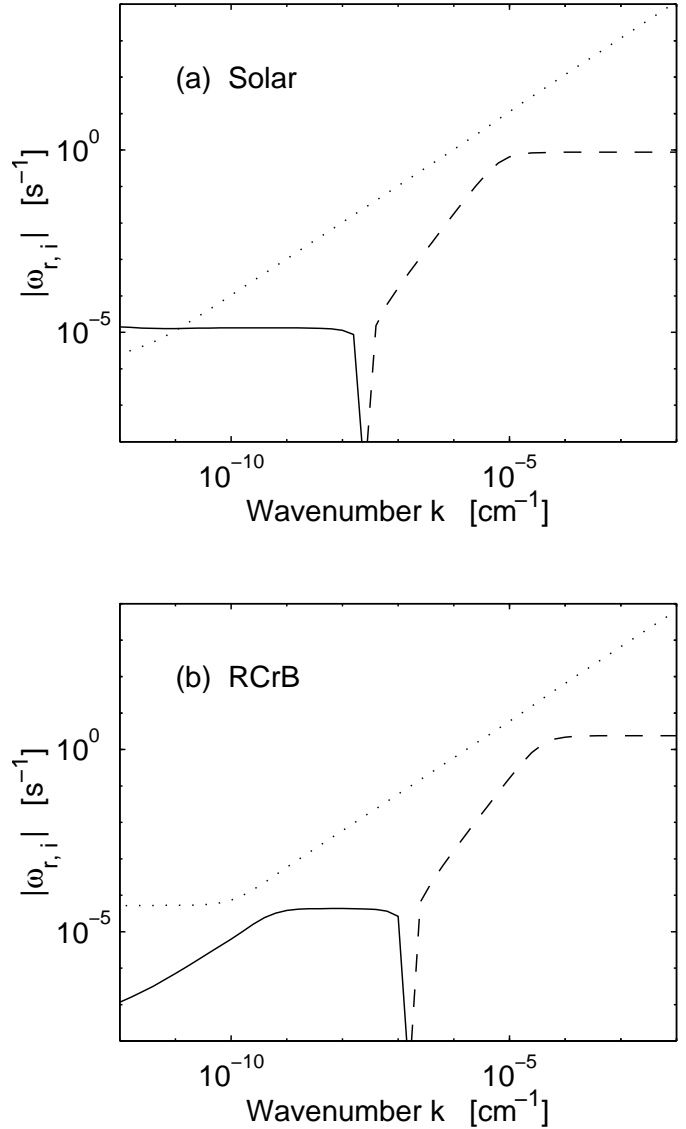,width=9cm}}
   \caption{The numerical solutions to the full set of
linearized equations for the gravity-acoustic wave modes for
{\bf a} solar and {\bf b} R\,CrB
compositions. The real parts are the dotted curves. 
The solid lines correspond
to amplification of the imaginary parts, while the
dashed curves represent damping. For wavenumbers with damping, 
the absolute magnitude of the imaginary part is shown}
   \label{f:growthrate}
\end{figure}

The calculated growth rates are
shown in Fig. \ref{f:growthrate}, which reveals amplification
for $k \la 2 \cdot 10^{-8}$\,cm$^{-1}$ (solar composition)
and $k \la 2 \cdot 10^{-7}$\,cm$^{-1}$ (R\,CrB composition). 
For large $k$, the perturbations are as expected damped 
due to the negligible optical depths of the disturbances
($\tau_\Lambda = 2\pi \kappa \rho / k$),
which make the sound waves little affected by the radiative forces.
Instead, the smoothing effects from radiation energy exchange
due to large gradients in the perturbed variables dominate.
The maximum growth rates are slightly smaller for a solar
composition than for a H-deficient composition: 
$1.3\cdot 10^{-5}$ respectively 
$4.3\cdot 10^{-5}$\,s$^{-1}$, 
which corresponds to an e-folding time on the order of the 
sound crossing time of the atmosphere.
For the largest wavenumbers with such growth rates, 
the real parts are significantly higher,
which implies that the 
sound waves are only slightly amplified per wavelength. 
For smaller $k$, however, the growth rate is comparable with 
the oscillation rate.
In fact, for the model with solar composition, the wave is
essentially non-propagating for the smallest $k$. However,
the use of a local analysis is not justified when the 
wavelengths of the perturbations are comparable with the
scale heights of the variations for the variables. 
For $k \ga 10^{-10}$\,cm$^{-1}$, the propagation speed is 
slightly less than the isothermal sound speed in both cases.

Thus, radiation-modified sound waves can be amplified in
late-type stars with significant radiative forces. 
However, the perturbations are not very strongly 
excited with minimum growth times $\simeq 10^5$\,s, 
contrary to the findings by Carlberg (1980) 
for hot stars, mainly as a result of the much smaller radiative
fluxes ($F \propto T_{\rm eff}^4$). 
For the instability to generate large amplification,
the sound waves must be reflected at the surface and pass
through the region with high radiative forces several times.
However, the inwards propagating modes are damped for the
same reason as the outgoing are amplified, with the magnitude
of the damping rate similar to the growth rate, which suggests
that the instability will still not be very efficient.
In order to estimate the behaviour of the instability 
in the global, non-linear
regime, a more sophisticated study 
than presented here must be carried out.

\subsection{The simplifying case of a 
isothermal, non-ionizing atmosphere}

In order to gain further theoretical
insight into which atmospheric properties contribute to
the instability presented above,
the equations can be simplified by assuming the 
unperturbed atmosphere to
be isothermal and neglecting ionization 
and scattering. 
Furthermore, only the limits of optically
thin and optically thick disturbances will be 
studied and hence the combined radiative energy and 
momentum equation will not be needed. 
The gas energy equation will thus instead be written as:

\begin{equation}
\rho\frac{{\rm D} e}{{\rm Dt}} -
\frac{P_{\rm gas}}{\rho} 
\frac{{\rm D\rho}}{{\rm Dt}} =  - {\cal L}
\end{equation}

\noindent where ${\cal L}$ denotes the net radiative cooling rate 
of the gas.
Depending on the optical thickness of the disturbance,
${\cal L}$ will be described by either Newtonian cooling
or equilibrium diffusion (e.g. Mihalas \& Weibel Mihalas 1984). 
The rate of work done by $g_{\rm rad}$ has been ignored.

The linearized continuity equation remains the same but
the linearized momentum equation reads as:

\begin{eqnarray}
\frac{\rho_1}{\rho_0}\left[
c_{\rm T}^2 ik + g_{\rm eff} - g_\rho\right] & + & \nonumber \\
+ \frac{T_1}{T_0}\left[
c_{\rm T}^2\left(ik - H_\rho^{-1}\right) - g_{\rm T}\right] +
v_1\left[-i\omega\right] & = & 0,
\end{eqnarray}

\noindent where $g_\rho = \rho_0 \partial g_{\rm rad}/\partial\rho$ 
and $g_{\rm T} = T_0 \partial g_{\rm rad}/\partial T$, while
the energy equation looks like:

\begin{equation}
\frac{\rho_1}{\rho_0}\left[-i\omega\right] +
\frac{T_1}{T_0}\left[\frac{i\omega}{\gamma -1} - 
\tilde{{\cal L}}\right] +
v_1\left[-H_\rho^{-1}\right] = 0.
\end{equation}

\noindent Here $\tilde{{\cal L}}$ represents the term arising
from the net cooling rate ${\cal L}$ and $\gamma$ the ratio
of specific heats.
For Newtonian cooling 

\begin{equation}
\tilde{{\cal L}} = 
\frac{16 \kappa_{\rm Ross} \sigma_{\rm R} T_0^4}{(\gamma -1)c_{\rm v}T}
\end{equation}

\noindent and in the diffusion regime

\begin{equation}
\tilde{{\cal L}} = 
\frac{16 \kappa_{\rm Ross} \sigma_{\rm R} T_0^4}{(\gamma -1)c_{\rm v}T}
\cdot \frac{k^2}{3\kappa_{\rm Ross}^2\rho^2}.
\end{equation}

\noindent $\tilde{{\cal L}}$ is 
the inverse of the radiative cooling time-scale
(e.g. Mihalas \& Weibel Mihalas 1984), where 
the factor $k^2/(3 \kappa_{\rm Ross}^2 \rho^2) \ll 1$ takes into
account the optical depth effects for optically thick 
conditions.
The simplified dispersion relation then reads:


\begin{eqnarray}
- i\omega^3 + \omega^2\tilde{{\cal L}}\left(\gamma-1\right) 
& + & \nonumber \\
+ i\omega\left[
\gamma k^2c_{\rm T}^2 + H_\rho^{-1}\left(g_{\rm eff} - g_\rho\right) 
\right] 
& + & \nonumber \\
- \omega k\left[g_\rho + \left(\gamma-1\right)g_{\rm T} - g_{\rm eff}
+ \gamma c_{\rm T}^2 H_\rho^{-1}\right] 
& + & \nonumber \\
-\tilde{{\cal L}}\left(\gamma-1\right)\left[
 k^2c_{\rm T}^2 + H_\rho^{-1}\left(g_{\rm eff} - 
g_\rho\right)\right] 
& + & \nonumber \\
- ik\tilde{{\cal L}}\left(\gamma-1\right)\left[
g_\rho - g_{\rm eff}
+ c_{\rm T}^2 H_\rho^{-1}\right] 
& = & 0.
\end{eqnarray}

\noindent For $k \ga 10^{-10}$\,cm$^{-1}$, 
$H_\rho^{-1}\left(g_{\rm eff} - 
g_\rho\right)$ is much smaller than $k^2c_{\rm T}^2$ 
for the two investigated models and can be ignored.

In situations of rapid radiative cooling 
($\tilde{{\cal L}} \gg \omega$), the dispersion relation
can be simplified to yield the solutions

\begin{equation}
\omega \simeq \pm \left[
kc_{\rm T}  
+\frac{i}{2c_{\rm T}}
\left\{g_\rho - g_{\rm eff} + H_\rho^{-1}c_{\rm T}^2\right\}
\right],
\end{equation}

\noindent i.e. isothermally propagating, radiation-modified
sound waves. In the limit of slow cooling
($\tilde{{\cal L}} \ll \omega$),
as expected adiabatic acoustic waves are recovered:

\begin{equation}
\omega \simeq \pm \left[
kc_{\rm s}  
+\frac{i}{2c_{\rm s}}
\left\{g_\rho + (\gamma - 1)g_{\rm T} - g_{\rm eff} + 
H_\rho^{-1}c_{\rm s}^2\right\}
\right],
\end{equation}

\noindent 
which travel with the adiabatic sound speed 
\mbox{$c_{\rm s}=(\partial P_{\rm gas}/\partial \rho)_{\rm s}=
\gamma P_{\rm gas}/\rho$}.
For $k \la 10^{-10}$\,cm$^{-1}$, the
propagation speed will be modified by the neglected terms
$H_\rho^{-1}\left(g_{\rm eff} - g_\rho\right)$, though
the growth/damping rates will remain the same. 

For both isothermal and adiabatic perturbations,
the sound waves are amplified by the action of 
the radiative acceleration, as long as the opacity increases
with temperature and density (assuming constant radiative flux):
 
\begin{eqnarray}
g_\rho & = & \frac{\sigma T_{\rm eff}^4}{c} \cdot \rho
\frac{\partial \kappa_{\rm F}}{\partial \rho} \\
g_{\rm T} & = & \frac{\sigma T_{\rm eff}^4}{c} \cdot T
\frac{\partial \kappa_{\rm F}}{\partial T}.
\end{eqnarray}

\noindent 
Furthermore, in a $P_{\rm gas}$-inversion $g_{\rm eff}$ is negative
and hence tends to excite an initial disturbance.
However, the normal momentum conservation
counteracts this due to propagation in a density stratified medium; 
the density inversion will tend to damp 
the amplitude for outwards propagating waves. 
In fact, with this simplified atmosphere
and the values in Table \ref{t:values}, 
in both models the last term dominates.
Hence, the amplitude of sound waves will decay  
as they propagate outwards with these
assumptions, despite the super-Eddington luminosities and
the increase in opacity upon compression. 
Thus, since the numerical results in Sect. \ref{s:radinstab} 
show amplification
for the case without these simplifying assumptions, the
effects of e.g. ionization and a temperature gradient 
tend to work together with the radiative force and its derivatives
to overcome the damping contribution from the 
density gradient.



\section{Conclusions and speculations}

It has been shown that 
H-deficient model atmospheres may
have radiative forces which exceed gravity, similar to
the case of some luminous H-rich stars
(Lamers \& Fitzpatrick 1988; Gustafsson \& Plez 1992).
Such super-Eddington luminosities manifest themselves as
$P_{\rm gas}$-inversions in hydrostatic atmospheres.
The inversions are, however, not an artifact of this assumption but
can also be present in dynamical atmospheres.
The inversions are not removed unless a very high mass
loss rate is present, and is neither necessarily removed
by convection if the large radiative forces occur 
close to the surface.
Thus, super-Eddington luminosities does {\it not} automatically initiate
a stellar wind.

The location of the H-deficient R\,CrB stars in the immediate proximity
of the computed opacity-modified Eddington limit 
(Fig. \ref{f:eddlimit}), is certainly suggestive of a 
connection between the enigmatic visual declines of the
stars and the Eddington limit. 
It is therefore proposed that instabilities as the R\,CrB stars 
encounter the Eddington limit during their evolution towards
higher $T_{\rm eff}$ are the unknown trigger mechanism 
for their famous variability.
The Eddington limit may thus be the ``smoking gun'' for the
declines: gas is ejected due to the high
radiative forces in the atmospheres, which then
cools rapidly first radiatively and later from adiabatic
expansion to reach sub-equilibrium temperatures and thus
possibly enable dust condensation (cf. Woitke et al. 1996
for the similar case of shocks instead induced by pulsations).
Such gas ejections may be observable as absorption components
in strong lines, as observed 
by Rao \& Lambert (1997) in R\,CrB at maximum light.
If a connection indeed exists, it could explain why
the EHe and HdC stars do not show R\,CrB-like variability
as a result of being located on the stable side of the
Eddington limit. 
It also points to an interesting similarity between the
R\,CrB stars and the LBVs.
Thus, the two types of stars known to be situated at the Eddington limit
both show eruptive behaviour,
which suggests a similar underlying physical
explanation for their variability.
 
A search for radiative and dynamical instabilities 
in the atmospheres has been carried out for both H-rich and
H-deficient late-type supergiants, but only 
partly successfully. Sound waves are found to
be amplified by the large radiative forces but the linear stability
analysis reveal only rather slow growth rates compared to for early-type
stars despite the super-Eddington luminosities. Such radiation-modified
sound waves may thus be partly responsible 
for the semi-regular pulsational variations of such supergiants,
but it is doubtful whether the instability is efficient enough
to eject gas clouds from the atmospheres as speculated above.
The atmospheres are also found to be close to dynamically
unstable, which might give rise to increased mass loss, 
as previously suggested for LBVs (Stothers \& Chin 1993)
and for the termination of the AGB phase 
(e.g. Paczy\'{n}ski \& Zi\'{o}{\l}kowski 1968; 
Wagenhuber \& Weiss 1994).
This does not, however, explain in what way the 
R\,CrB stars are special compared to other similar stars. 
In fact, the H-rich models seem to be more vulnerable to 
such instabilities, which suggests that dynamical instabilities
are not the answer to the declines of the R\,CrB stars. 
Another possibility is of course that the same dynamical
instabilities occur in both types of stars 
but that the H-deficient and C-rich environment
of the R\,CrB stars greatly favours dust condensation, and 
thus the observational manifestation, compared to
for H-rich compositions. 

One can also imagine that the Eddington limit is still
responsible for the behaviour of the R\,CrB stars 
though in a more indirect way. 
Langer (1997a,b) has shown that the effect of rotation 
coupled with strong radiative forces may lead to increased
mass loss in the equatorial plane, and it is possible 
that such a mechanism is at work in these supergiants,
despite the presumably relatively small rotational velocities. 
The difference in observed variability between the R\,CrB stars 
and the HdC and EHe stars could then be
the result of viewing angle.
There are some observational evidence for such bipolarity
(e.g. Rao \& Lambert 1993; Clayton et al. 1997)
Violent instabilities due to strange mode pulsations 
leading to dramatically increased mass loss is
another possibility,
which is partly related to the Eddington
limit, since both have their origin in ionization zones, as do 
the dynamical instabilities discussed above. 
Whether any of these instabilities, or a combination thereof, could 
help explain the variability of the R\,CrB stars and the LBVs
certainly deserves further investigation.

Finally, it should be remembered that the 
stability analyses presented here
assume the unperturbed model to be a realistic description
of the stellar atmosphere.
At least for the R\,CrB stars, 
there are strong indications that this is 
in fact not the case 
(Gustafsson \& Asplund 1996; Asplund et al. 1997a,b,c;
Lambert et al. 1997).
If the supergiant atmospheres
are indeed distinctly different
from the predictions of standard models, 
the efficiency of 
radiative instabilities may have been underestimated.
One can suspect that 
models based on the normal assumptions, such as the mixing length
theory for convection, 
near the Eddington limit may be very unsuitable.
Clearly, radiative hydrodynamical simulations of such atmospheres
would be valuable, also since they 
would shed further light on possible instabilities.

\begin{acknowledgements}
The constructive suggestions by F.P. Pijpers and A. Weiss have 
significantly improved the paper. Discussions with 
W. Glatzel, B. Gustafsson, 
L.-A. Willson and P. Woitke have been very helpful. The author is grateful 
to L. Achmad, H.J.G.L.M. Lamers, N. Langer and S. Owocki 
for communicating their work 
prior to publication.
\end{acknowledgements}

\end{document}